\documentclass[twocolumn]{aastex62}
\usepackage[english]{babel}
\usepackage{amsmath}

\usepackage{xspace}

\newcommand{\jybm}{Jy\,beam$^{-1}$\xspace}
\newcommand{\kms}{km\,s$^{-1}$\xspace}

\shorttitle{Resolving the disk of HD\,100546}
\shortauthors{J.E. Pineda et al.}

\begin{document}

\title{High-Resolution ALMA Observations of HD\,100546: 
  Asymmetric Circumstellar Ring, and Circumplanetary Disk Upper Limits}

\author[0000-0002-3972-1978]{Jaime E. Pineda}
\affiliation{Max-Planck-Institut f\"ur extraterrestrische Physik, Giessenbachstrasse 1, 85748 Garching, Germany}
\email{jpineda@mpe.mpg.de}

\author[0000-0001-8442-4043]{Judit Szul\'agyi}
\affiliation{Center for Theoretical Astrophysics and Cosmology, Institute for Computational Science, University of Zurich, Winterthurestrasse 190, CH-8057 Zurich, Switzerland}
\affiliation{Institute for Particle Physics and Astrophysics, ETH Zurich, Wolfgang Pauli Strasse 27, CH-8093 Zurich, Switzerland}

\author[0000-0003-3829-7412]{Sascha P.\,Quanz}
\affiliation{Institute for Particle Physics and Astrophysics, ETH Zurich, Wolfgang Pauli Strasse 27, CH-8093 Zurich, Switzerland}
\affiliation{National Center of Competence in Research "PlanetS" (\url{http://nccr-planets.ch})}

\author[0000-0001-7591-1907]{Ewine F. van Dishoeck}
\affiliation{Leiden Observatory, Leiden University, PO Box 9513, NL-2300 RA Leiden, the Netherlands}
\affiliation{Max-Planck-Institut f\"ur extraterrestrische Physik, Giessenbachstrasse 1, 85748 Garching, Germany}

\author[0000-0002-4266-0643]{Antonio Garufi}
\affiliation{Universidad Aut\'ononoma de Madrid, Dpto. F{\'\i}sica Te\'orica, M\'odulo 15, Facultad de Ciencias, E-28049 Madrid, Spain}
\affiliation{INAF/Osservatorio Astrofisico of Arcetri, Largo E. Fermi, 5, I-50125 Firenze, Italy}

\author[0000-0002-3984-9496]{Farzana Meru}
\affiliation{Institute of Astronomy, Madingley Road, Cambridge, CB3 0HA, UK}

\author[0000-0002-1078-9493]{Gijs D. Mulders}
\affiliation{Lunar and Planetary Laboratory, The University of Arizona, 1629 E. University Blvd., Tucson, AZ 85721, USA}

\author[0000-0003-1859-3070]{Leonardo Testi}
\affiliation{European Southern Observatory, Karl-Schwarzschild-Str. 2, D-85748 Garching bei M\"unchen, Germany}
\affiliation{INAF/Osservatorio Astrofisico of Arcetri, Largo E. Fermi, 5, I-50125 Firenze, Italy}

\author[0000-0003-1227-3084]{Michael R. Meyer}
\affiliation{Institute for Astronomy, ETH Zurich, Wolfgang-Pauli-Strasse 27, 8093 Zurich, Switzerland}
\affiliation{Department of Astronomy, University of Michigan, 500 Church Street, Ann Arbor, MI 48109, USA}

\author[0000-0003-2911-0898]{Maddalena Reggiani}
\affiliation{D\'epartement d'Astrophysique, G\'eophysique et Oc\'eanographie, Universit\'e de Li\`ege, 17 All\'ee du Six Ao\^ut, 4000, Li\`ege, Belgium}

\begin{abstract}
We present long baseline Atacama Large Millimeter/submillimeter Array (ALMA) observations of the 870$\,\mu$m dust continuum emission and CO (3--2) from the protoplanetary disk around the Herbig Ae/Be star HD\,100546, which is one of the few systems claimed to have two young embedded planets. 
These observations achieve a resolution of 4\,au (3.8\,mas), an rms noise of 66\,$\mu$\jybm, and reveal an asymmetric ring between $\sim$20--40 au with largely optically thin dust continuum emission.
This ring is well fit by two concentric and overlapping Gaussian rings of different widths and a Vortex.
In addition, an unresolved component is detected at a position consistent with the central star, which  may trace the central inner disk ($<$2\,au in radius).
We report a lack of compact continuum emission at the positions of both claimed protoplanets.
We use this result to constrain the circumplanetary disk (CPD) mass and size of 1.44\,M$_\earth$ and 0.44\,au 
 in the optically thin and thick regime, respectively, for the case of the
previously directly imaged protoplanet candidate at $\sim$55\,au (HD100546 b). We compare these empirical CPD constraints to previous numerical simulations. This suggests that HD100546 b is inconsistent with several planet accretion 
models, while gas-starved models are also still compatible. 
We estimate the planetary mass as 1.65\,M$_J$ by using the relation between planet, circumstellar, and circumplanetary masses derived from numerical simulations.
Finally, the CO integrated intensity map shows a possible spiral arm feature that would match the 
spiral features identified in Near-Infrared scattered light polarized emission, which suggests a real spiral feature in the disk surface 
that needs to be confirmed with further observations.
\end{abstract}

\keywords{stars: pre-main sequence --- stars: formation --- protoplanetary disks --- planet-disk interactions ---  stars: individual (HD\,100546) --- Techniques: interferometric}

\section{Introduction}

Gas and dust rich disks around young stars are the birthplace of new planetary systems. 
However, we still lack observational data showing under which physical and chemical conditions gas giant planet formation takes place. Radial velocity (RV) exoplanet surveys have shown that 6-7\% of solar type stars host gas giant planets in the inner few au, and that the occurrence rate of these planets increases with stellar mass \citep{Cumming_2008,Johnson_2010,Wittenmyer2016}. Combining RV data with high contrast imaging follow-up, \citet{Bryan2016} suggest that the total occurrence rate of companions with masses from 1-20 M$_{\rm Jupiter}$ and separations from 5-20~au could be as high as $\approx$50\%. 
In contrast, high-contrast direct imaging surveys reveal that beyond 50~au massive giant planets are 
very rare \citep[e.g.,][]{Lafreniere2007,Chauvin2010,Heinze2010, Rameau2013b, Biller2013, Nielsen2013, Wahhaj2013, Chauvin2015, Meshkat2015, Reggiani_2016}.  
However, planets of a few $M_J$ have been directly imaged around a few stars at orbital separations between 10 and 70~au \citep[e.g., HR8799, $\beta$ Pictoris, HD95086, 51 Eri, HIP65426;][]{Marois_2008-HR8799_Planets,Lagrange_2010-betaPic_planet,Rameau2013a,Macintosh_2015,Chauvin2017}. 

On the theoretical side, there are two main theories for gas giant planet formation: the core accretion (CA) paradigm \citep[e.g.,][]{Pollack_1996-Core_Accretion}
and the gravitational instability (GI) theory \citep[e.g.,][]{Boss_2001}. The former one, which is based on the initial growth of solids to eventually form the cores of gas giant planets, has recently been modified to allow for a more rapid accretion of cm and dm sized particles \citep[pebble accretion,][]{Lambrechts2012}. It is unknown which of the mechanisms is responsible for the observed giant planet population or whether all of them contributed in different amounts 
\citep[see][for a recent review]{Helled2014-PPIV}.

To address these fundamental issues it is crucial to detect and study young giant planets in their formation phase, when they are still embedded in their natal environment. 
An elegant way to investigate the formation mechanism is to study the properties of the circumplanetary disk (CPD) that 
surrounds the young planet and transports material from the circumstellar disk (CSD) onto the forming object. CPD properties have been shown to be strongly dependent on the planet formation mechanism \citep{Szulagyi2017_CA_DI}. 
While analytic and numerical simulations generally agree that, irrespective of the formation mechanism, the CPD radius should be a fraction of the planet's Hill radius \citep{Quillen_1998-ProtoJovian_Outflows,Ayliffe_2012-protoplanet_formation,Shabram_2013-Disk_Planet_Simulation}, their masses and temperatures are expected to be significantly different, with GI leading to more massive but colder CPDs compared to CA \citep{Szulagyi2017_CA_DI}. 
Hence, the direct detection of emission from CPDs, shedding light on their size and mass, would be a 
major step in understanding how gas giant planets are formed.

Up to now, a few systems show direct evidence, based in high-contrast imaging observations, 
of candidate gas giant planets that are still in their formation phase: HD100546, which is subject of this paper (see details on the system below), LkCa15 \citep{KrausIreland_2012-LkCa15, Sallum2015}, HD169142 \citep{Reggiani2014,Biller2014}, MWC 758 \citep{Reggiani2018_MWC758} and PDS70 \citep{Keppler2018}. 
\cite{Isella2014-CPD} searched for CPD dust continuum emission in LkCa15 with the VLA, but did not succeed. 
For HD100546 and HD169142, the very red near-infrared (NIR) colors of the companion candidates  are inconsistent with pure photospheric emission of young gas giant planets, which led to the suggestion that the observed fluxes are a superposition of emission from a young planet and an additional CPD \citep{Quanz2015-HD100546b,Reggiani2014}. More recently, for HD100546 b, the emission from the CPD has been predicted to be $800\mu$Jy at 870$\mu$m \citep{Zhu2016_CPD}.
Here, we present an analysis of new ALMA Cycle 3 observations of the 
870\,$\mu$m dust continuum emission of HD100546 reaching an rms noise of 66\,$\mu$\jybm and with high enough angular resolution to separate the CPD and CSD. 

\section{The HD~100546 system\label{sec-source}}
HD\,100546 is a Herbig Ae/Be star 
located at a distance of 110.02$\pm$0.62\,pc \citep{Gaia2018_DR2}. 
%
% GAIA DR2 
% parallax=9.08919838108378 mas,   sigma_parallax=0.05088846005791359 mas
%
The transition disk around this star has a cavity (in dust and molecular gas) between $\sim$1--14\,au 
\citep[e.g.,][]{bouwman2003,grady2005,benisty2010,quanz2011,Mulders_2013-HD100546_Companion_Mass,Panic_2014-HD100546_Disk_Asymmetry,liskowsky2012,brittain2009,vanderplas2009,Liu_2003-HD100546_Resolved_Disk,Sissa2018}.
The major axis is located at 145.14$\pm$0.04 east of north \citep{Pineda2014}.  %
The presence of a companion (HD100546 c) inside this cavity was suggested by various studies based on both indirect and direct evidence \citep[e.g.,][]{bouwman2003,Acke_2006-VLT_HD100546_Rotation_Companion,tatulli2011,brittain2013,Mulders_2013-HD100546_Companion_Mass}. 
However, \cite{Fedele2015} put forward an explanation that the spectroastrometric signature seen in the rovibrational CO emission lines \citep{brittain2013} does not require a planet, and 
\cite{Follette2017} claim that the uncertain direct imaging detection from \cite{Currie2015-HD100546} is caused by aggressive data processing.
An additional protoplanet candidate (HD100546 b) was identified further out 
in the outer disk ($\sim$50--60\,au separation from the central star) using high-contrast direct imaging observations 
\citep{Quanz_2013-NACO_HD100546,Currie2014-HD100546,Quanz2015-HD100546b,Currie2017}.  
However, this detection was called into question in particular because no accretion features were detected \citep{Rameau2017}.% 

The current best dust continuum data published for HD\,100546 are from ALMA C0 at 870 $\,\mu$m  \citep{Pineda2014,Walsh2014} with 0.6\,$\arcsec$ resolution, and
from the Australia Telescope Compact Array (ATCA) 
at 7\,mm with an angular resolution of 0.15\,$\arcsec$ \citep{Wright2015}.
Both analyses of the 
ALMA C0 data \citep{Pineda2014,Walsh2014} identified (in the uv-space) a ring-like structure 
of the dust emission that is more compact than the gas,
while \cite{Walsh2014} also identified a second fainter ring further out. 
However, the main discrepancy between these two works is the 
claim of an asymmetry in the dust continuum emission based on the residuals from the comparison of the best fit model with the data by \cite{Pineda2014}, 
while \cite{Walsh2014} claim that their emission is symmetric based on the analysis of the interferometric visibilities. 
On the other hand, \cite{Wright2015} claim an asymmetry at 7 and 3\,mm in the images, 
but in the opposite direction as reported by \cite{Pineda2014}. 
The presence of asymmetries have been revealed in protostellar disks \citep[e.g.,][]{vanderMarel_2013-OphIRS48_ALMA,Casassus_2015-HD142527,Kraus2017-V1247,Perez_2014-ALMA_Asymmetry}, 
which have implications on the planet formation mechanism at play and their related timescales \citep[e.g.,][]{Lyra_2013-Vortex,Mittal_2015-Horseshoe}. 
Therefore, and in order to search for direct evidence for CPDs, data with higher angular resolution and image fidelity were needed to settle this issue.

\subsection{Updated stellar parameters\label{sec-stelar-par}}
The most up-to-date and accurate distance estimate 
(110.02\,pc from GAIA DR2) to the star is larger than the previously 
derived (97\,pc from Hipparcos), which was used to estimate the 
stellar parameters. 
and therefore we refine the stellar parameters based on d=110.02\,pc. 
We adopt a PHOENIX model of the stellar photosphere \citep{Hauschildt1999:Paper1} 
with $T_{\rm eff}=9,800$\,K \citep{Fairlamb2015} and $\log(g)=$-4.0, 
then it is scaled to the GAIA DR2 distance and to the de-reddened 
($A_{\rm V}=0.1$\,mag) $V$-band photometry. 
The integrated luminosity $L_*$ is calculated from the model, 
which combined to the aforementioned $T_{\rm eff}$ are compared 
to the Pre-Main Sequence (PMS) stellar tracks by 
\cite{Siess2000}. 
We employed the tracks with depleted abundance of $Z$, 
because the source is depleted in refractories elements in its atmosphere \citep{Folsom2012}. 
This procedure yields a stellar mass and age of
$M_*=2.2\pm0.2$\,M$_\odot$ and 
$t=4.8^{+2.0}_{-1.1}$~Myr, respectively. 
The reported uncertainties are obtained by propagating the
uncertainties on the distance, $A_{\rm V}$ ($\pm0.1$), and $T_{\rm eff}$ (a conservative $\pm400$\,K).

%%%%%%%%%%%%%%%%%%%%%%%%%%%%%%%%%%%%%%%%
%%%%%%%%%%%%%%%%%%%%%%%%%%%%%%%%%%%%%%%%
\begin{figure}[ht]
\centering
\includegraphics[height=0.35\textwidth]{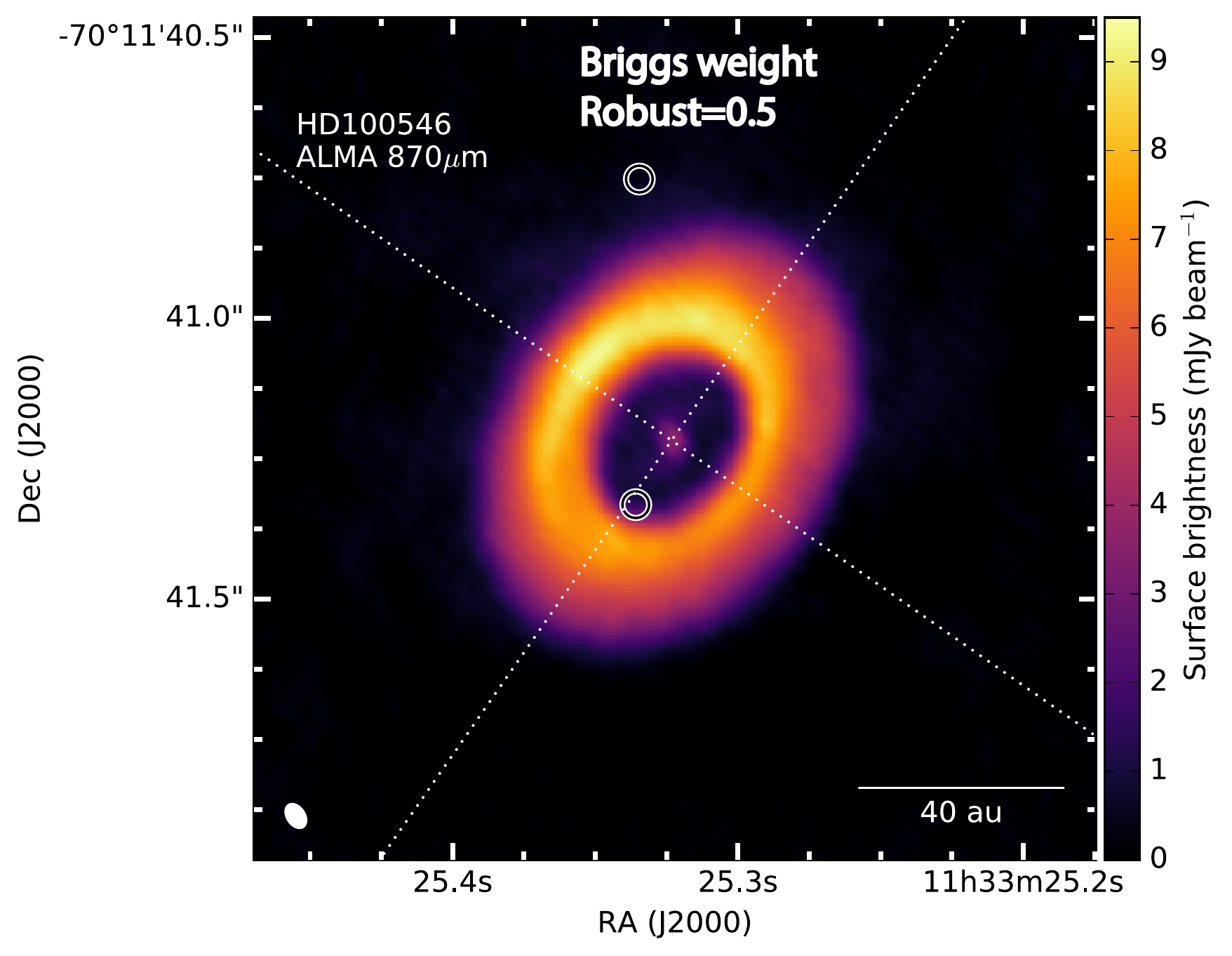}
\caption{Synthesized image of the 870\,$\mu$m continuum emission from the HD~100546 disk using 
Briggs robust weight of $0.5$, 
with an rms of 66\,$\mu$\jybm and a beam of 47$\times$31\,mas. 
Beam size and scale bar are shown in bottom left and right corners, respectively.. 
The markers show the positions of the claimed planets in the system. 
The dotted lines show the direction of the disk major and minor axes.
\label{fig-cont-maps}}
\end{figure}

\section{Data\label{sec-data}}

HD\,100546 was observed on 2015 December 2nd with ALMA using Band 7 receivers under 
project 2015.1.00806.S (PI: J.E. Pineda). 
The array configuration included 36 antennas with baselines between 17 and 10800\,m, 
but with insufficient short baselines ($<$100\,m) to properly recover emission at scales larger than $\approx 1$\,\arcsec.
The observations cycled through HD\,100546 and quasar J1147$-$6753 with a cycle time of $\sim 1$ minute. 
The bright quasar J1427$-$4206 was used as bandpass calibrator, while J1107$-$4449 was used to set the flux amplitude. 
The standard flagging and calibration was done using CASA 4.5.1 \citep{McMullin2007}, 
while imaging was done using CASA 4.7 and multiscale clean. 
Self-calibration was performed with the shortest phase and amplitude cycle of 10 and 60 seconds, respectively.
The 870\,$\mu$m continuum was obtained from line free channels and 
imaged using natural weighting to achieve an 
angular resolution of 0.056$\arcsec \times$0.041$\arcsec$ (PA=26.9$\degr$), 
with an rms noise of 86\,$\mu$\jybm, as estimated from emission free regions.
Similarly, we imaged the continuum using a Briggs weight of $0.5$, which results in an 
angular resolution of 0.047$\arcsec \times$0.031$\arcsec$ (PA=33.9$\degr$), 
and an rms noise of 66\,$\mu$\jybm, as estimated from emission free regions.
Figure~\ref{fig-cont-maps} shows the map using Briggs weight.

\begin{deluxetable}{lcc}[ht]
\tablecolumns{3}
\tablecaption{Obsevational parameters\label{table-observation}}
\tablehead{
\colhead{Parameter} & \colhead{Unit} & \colhead{Value} 
}
\startdata
\cutinhead{Phase Center}
R.A. & (hh:mm:ss.sss) & 11:33:25.318652\\
Dec. & (dd:mm:ss.sss) & -70:11:41.23173\\
\cutinhead{Continuum (Briggs weighting, Robust=0.5)}
Wavelength &  ($\mu$m) & 870\\
Peak Flux &  (\jybm) & 9.27\\
Total Flux &  (Jy) & 1.27\\
Beam Major axis &  (arcsec) & 0.047\\
Beam Minor axis &  (arcsec) & 0.031\\
Beam PA &  ($\degr$) & 33.9\\
rms &  ($\mu$\jybm) & 66\\
\cutinhead{CO (3--2) (natural weighting)}
Beam Major axis &  (arcsec) & 0.059\\
Beam Minor axis &  (arcsec) & 0.044\\
Beam PA &  ($\degr$) & 18.22\\
channel width & (\kms) & 0.209\\
rms &  (m\jybm\,channel$^{-1}$) & 5.8\\
\enddata
\end{deluxetable}

The total integrated flux of the image is 1.27 and 1.29\,Jy for the robust and naturally weighted images, respectively. 
This flux is consistent with the total flux measured in the ALMA C0 data.

We use the naturally weighted image when studying the circumstellar disk structure, while we use the image with robust weighting when investigating the existence of CPDs.

The CO~(3--2) data cube is obtained from the continuum subtracted visibilities 
resulting from using the task \verb+uvcontsub+, after applying the 
self-calibration solutions obtained from the continuum. 
The imaging is done using multiscale clean with natural weighting, which produced 
a beam size of 0.059$\arcsec \times$0.044$\arcsec$ (PA=18.22$\degr$). 
Natural weighting is used, because it provides the highest sensitivity to spectral line observations.
We estimate the rms in the spectral cube using the line-free channels as
5.8\,m\jybm\,per channel, 
with a channel width of 0.209\,\kms and a spectral resolution of two channels.
In this case, the clean mask is defined for each channel around the bright 
emission, however, still some imaging artifacts are present due to the 
missing short spacings. 

\begin{figure}[ht]
\centering
\includegraphics[width=0.49\textwidth]{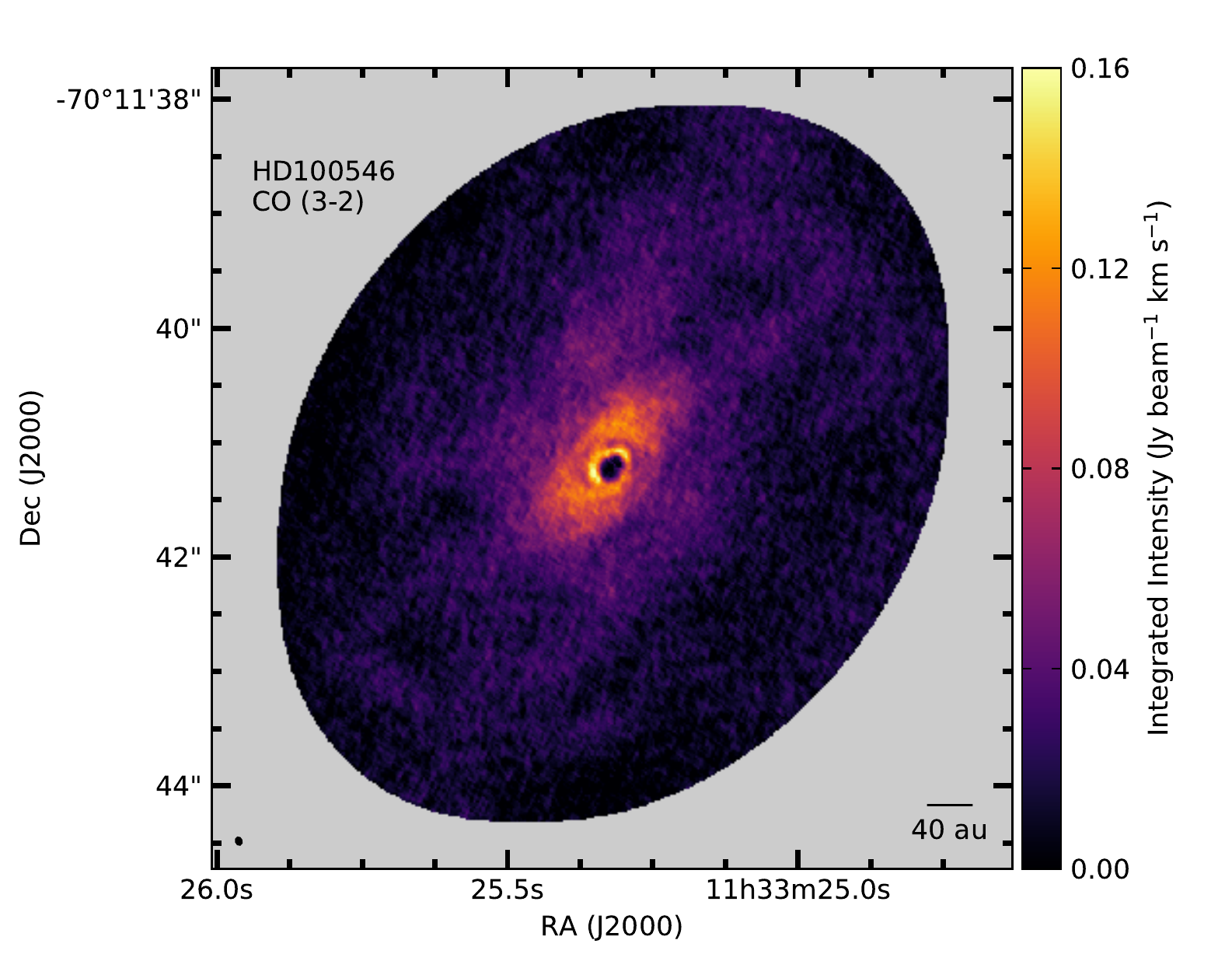}
\caption{Integrated intensity map of high-resolution CO~(3--2) emission 
for HD~100546 disk calculated using the Keplerian velocity mask.
The field-of-view shown is larger than continuum image show in 
Fig.~\ref{fig-cont-maps}.
Beam size and scale bar are shown in bottom left and right corners, respectively.
\label{fig-CO-TdV}}
\end{figure}

We use a Keplerian mask to calculate the moment maps, 
which is a similar to \cite{Friesen2017-GAS,Bergner2018,Calcutt2018} where the region used in the calculation 
is limited to voxels (3D pixels) close to the emission. 
In order to create the mask, we calculate the predicted Keplerian velocity at each pixel, 
where we assume the stellar parameters derived in Sec.~\ref{sec-stelar-par} and the 
disk parameters inclination and position angle derived from the continuum fit (see Table~\ref{table-vortex}) and a 
disk radius of 352\,au to match the extension of the CO emission as seen in the Cycle~0 observations.
The velocity field is then convolved with the same beam of the CO observations.
Finally, only voxels that are within 2\,\kms (similar to the linewidth in the inner part of the disk, $<$150\,au)
from predicted Keplerian velocity are kept in the final mask. 
The resulting integrated intensity map using the described mask is shown in Fig.~\ref{fig-CO-TdV}.

The total flux of the integrated intensity CO cube is 190\,Jy\,\kms, which is in excellent agreement with 
the total integrated intensity CO reported by \cite{Panic_2009-Disk_Models} using APEX of 191\,Jy\,\kms.

\section{Results\label{sec-results}}
\subsection{Maps and brightness profile\label{sec-Tprof}}

The maps shown in Figure~\ref{fig-cont-maps} reveal a bright ring between 20 and 40 au with a significant flux asymmetry, and an additional inner disk coincident with the position of the star. 
The inner disk is unresolved ($<2$\,au in radius) with a peak flux of $2.60 \pm 0.85$\,m\jybm. 
Between the inner and outer disk there is a dark annulus with an average brightness of $\sim 1$\,m\jybm, which is about 8$\times$ fainter than the (faintest section of the) 
central annulus of the ring emission.

\begin{figure}[ht]
\plotone{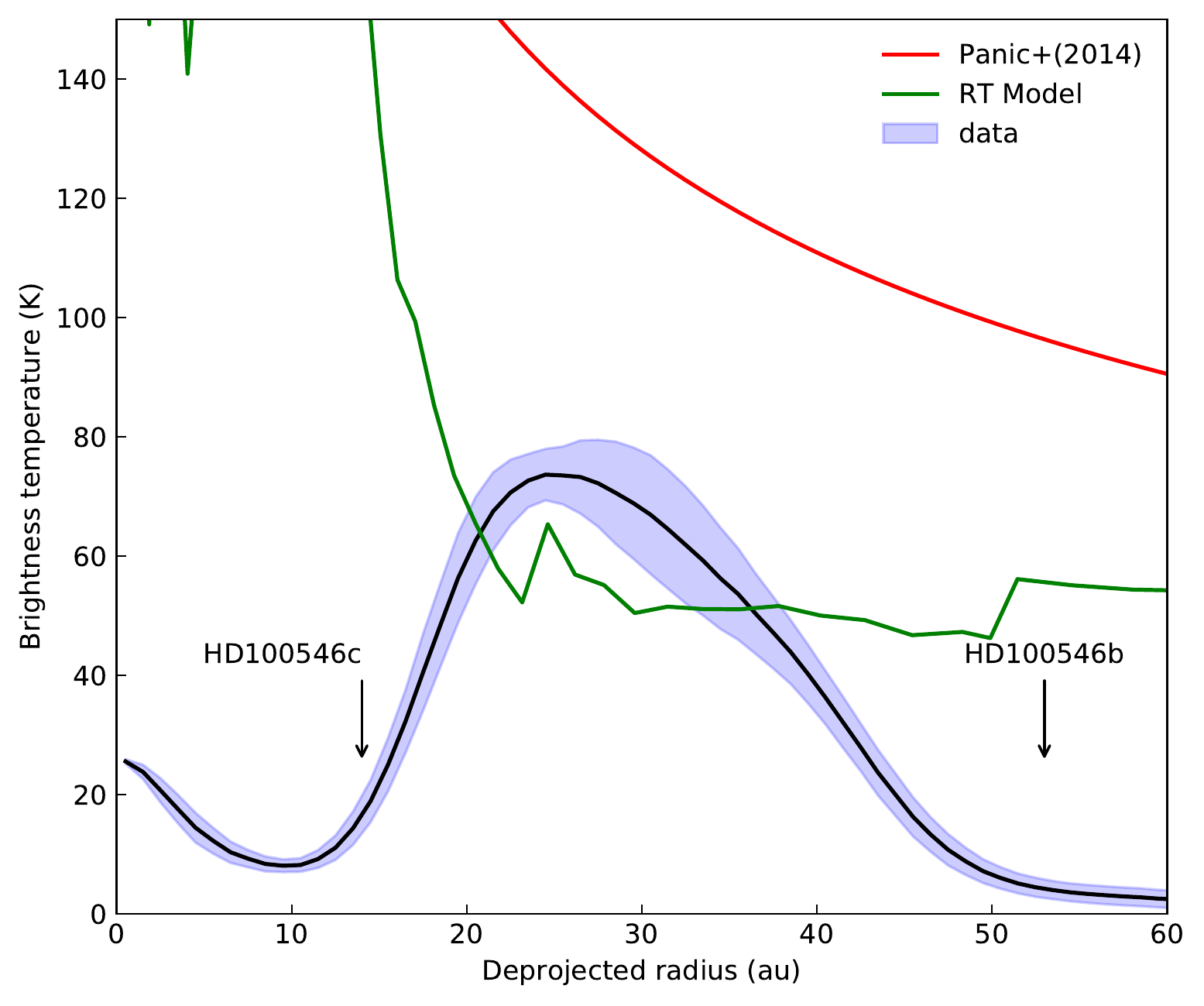}
\caption{Azimuthally averaged brightness temperature profile (black line). The shaded area shows the local standard deviation of the measurements. The average is done on the de-projected disk geometry.
The red curve is the temperature profile 
used by \cite{Panic_2014-HD100546_Disk_Asymmetry},
and the green curve is the temperature profile of the best radiative transfer model from \cite{Pineda2014}. 
The vertical arrows mark the expected position of the two planet candidates at 14 and 53~au.
\label{fig-T-r}}
\end{figure}

In Figure~\ref{fig-T-r} we compare the azimuthally averaged brightness temperature of the disk emission,  
for which we have calculated the deprojected radius using the position angle and disk inclination parameters obtained by \cite{Pineda2014}.
The same figure includes the parametric disk temperature profile from \cite{Panic_2014-HD100546_Disk_Asymmetry} and 
the temperature profile of the millimeter sized grains from the radiative transfer model from \cite{Pineda2014} (similar values of the mid-plane dust temperature at 50~au ($\approx$60~K) were found  by \cite{Bruderer_2012-HD100546}). 
The figure shows that at every position in the disk the parametric disk temperature from \cite{Panic_2014-HD100546_Disk_Asymmetry} 
is much higher than the observed values.
However, the more detailed radiative transfer model reveals lower temperatures for the millimeter sized particles, 
with an average temperature of $T_{d,mean}=$53~K between 20 and 50~au. 
We use this average disk dust temperature, $T_{d,mean}$, as a best estimate of the disk emission in the following sections.
Therefore the disk dust continuum emission is optically thin at the positions of the young planet candidates, 
while the central part of the ring might be optically thick.

\subsection{Parametric model}

We model the emission with a simple parametric model that includes a 2 Gaussian rings, a central compact source, and a vortex (to account for flux asymmetries), as follows:
\begin{eqnarray}
F(r, r_1,\theta)&=& F_{r1} e^{-(r-r_{r1})^2/2 \sigma_{r1}^2} \nonumber\\
&+& F_{r2} e^{-(r-r_{r2})^2/2 \sigma_{r2}^2} \nonumber\\
&+& F_g\, e^{-r_p^2/2 \sigma_g^2}/(2\pi\, \sigma_g^2) \nonumber\\
&+& F_V\, e^{-(r-r_v)^2/(2\,\sigma_v^2)} e^{-(\theta-\theta_v)^2/(2*\sigma_\theta^2)}
\end{eqnarray}
where $r$ and $r_r$ are the radii calculated at the center of the ring and point source, respectively, and taking into account the inclination angle with respect to the sky (assumed the same for both coordinate systems). 
The first two elements in the model attempts to reproduce the main disk ring-like emission (which is not well reproduced by a single Gaussian profile) and are concentric, the third one describes the central unresolved source, while the fourth element describes a possible vortex.

\begin{figure*}[ht]
\includegraphics[width=0.3\textwidth]{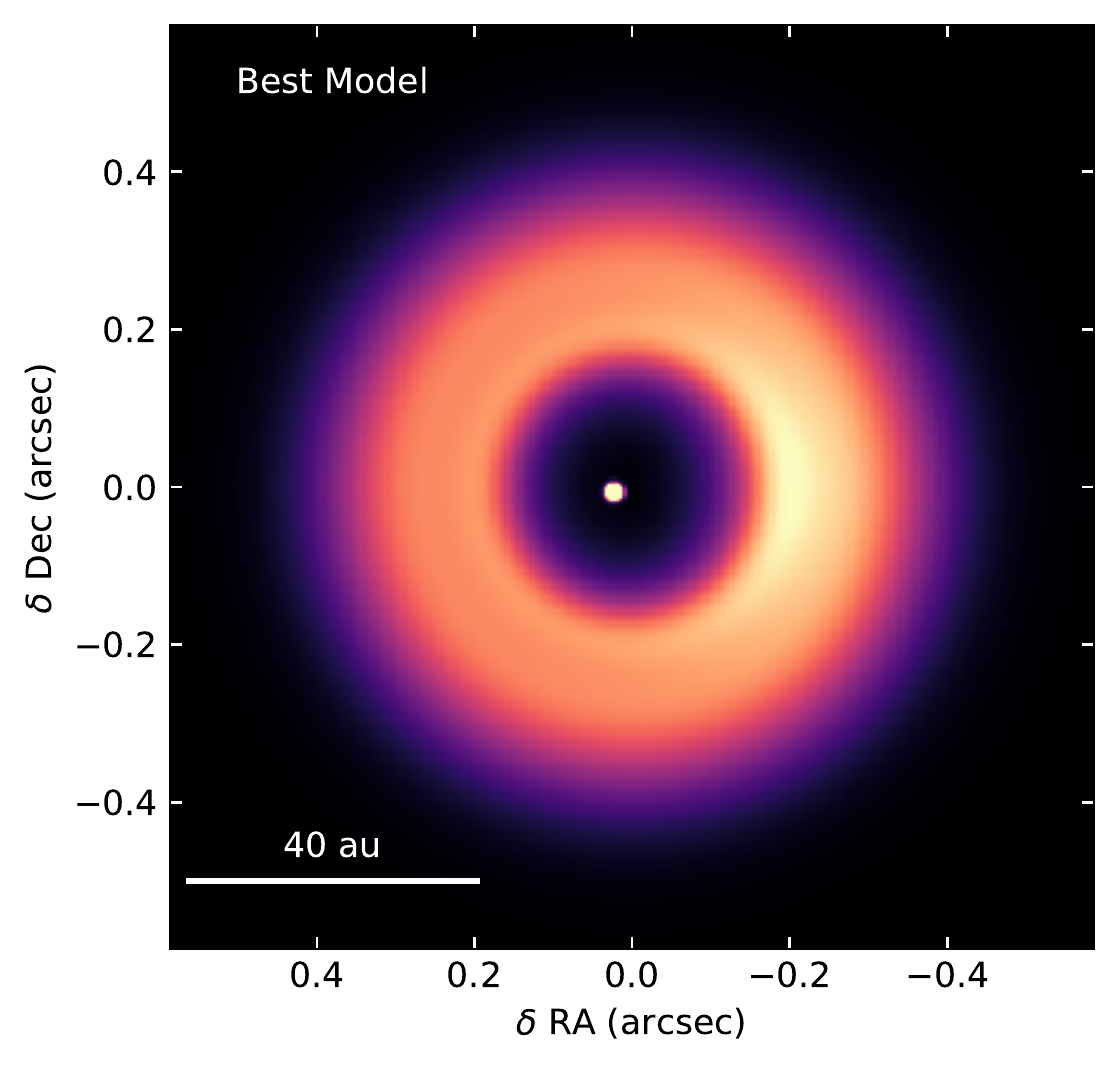}
\includegraphics[width=0.6\textwidth]{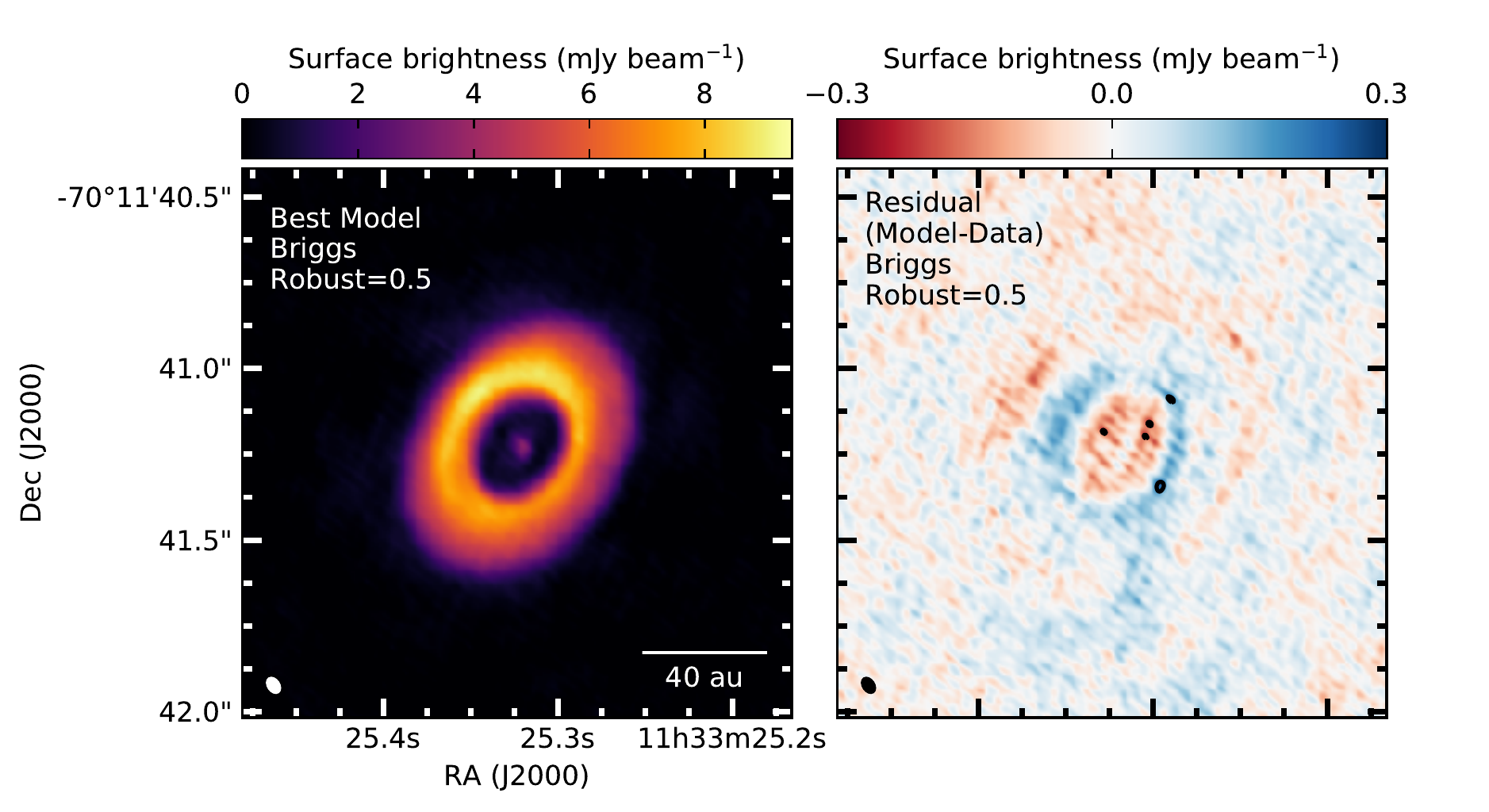}
\caption{\label{fig-vortex-model}
Left: the best fit parametric model in deprojected coordinates. 
Middle and right panel: best fit model and residual (model $-$ data) images for Briggs weighting of $0.5$.
Beam and scale bar are shown in the bottom left and right corners, respectively. 
The color stretch in the middle panel is the same as used in Fig.~\ref{fig-cont-maps}, and it shows the good level of 
agreement of the model with the data.
The contours on the right panel correspond to 3-$\sigma$ contours, where $\sigma$ is the reported noise level on the observed map.}
\end{figure*}

We use \verb+GALARIO+ \citep{Tazzari_2017-GALARIO} to sample the model image on the same visibilities 
as the observations. 
The $\chi^2$ is then calculated using the sampled visibilities, and then minimized in Python to find the optimal model. 
Also, we use the built-in options in \verb+GALARIO+ to perform 2D translations in the plane of the sky ($\delta$ RA, $\delta$ Dec) and rotation ($\delta$ PA) of the parametric model. 
The best fit model  in de-projected coordinates, observed model, and the residuals are shown in Fig.~\ref{fig-vortex-model}, 
while the best parameter values 
are listed in Table~\ref{table-vortex}.
The observed model and residuals are produced by sampling the same visibilities as the data, and then performing the imaging in CASA.

\begin{deluxetable*}{cccC}[ht]
\tablecolumns{4}
\tablecaption{Best fit parameters
\label{table-vortex}}
\tablehead{\colhead{Parameter} & \colhead{Unit} & \colhead{Meaning} & \colhead{Value}
}
\startdata
\cutinhead{System Geometry}
$\delta RA$ & ($10^{-3}$ arcsec) & RA Offset from phase center & 20.6\\
$\delta Dec$ & ($10^{-3}$ arcsec)& Dec Offset from phase center & 12.2\\
PA & (degrees) & Paralactic angle of the model\tablenotemark{1}& 139.1\\
incl & (degrees) & Inclination of the model\tablenotemark{2} &42.46\\
\cutinhead{Ring \#1}
$F_r$ & (Jy arcsec$^{-2}$) & Ring peak surface brightness & 1.50 \\
$r_v$ & (arcsec) & Ring radius & 0.186\\
                 & (au) & Ring radius & 20.5\\
$\sigma_v$ & (arcsec) & Ring width & 0.0303\\
                          & (au) & Ring width & 3.33\\
\cutinhead{Ring \#2}
$F_{r2}$ & (Jy arcsec$^{-2}$) & Ring peak surface brightness & 4.38\\
$r_{v2}$ & (arcsec) & Ring radius & 0.270\\
                     & (au) & Ring radius & 29.7\\
$\sigma_{v2}$ & (arcsec) & Ring width & 0.0919\\
                              & (au) & Ring width & 10.1\\
\cutinhead{Vortex}
$F_v$ & (Jy arcsec$^{-2}$) & Vortex peak surface brightness & 1.31\\
$r_v$ & (arcsec) & Vortex radius & 0.198\\
                & (au) & Vortex radius & 21.8\\
$\sigma_v$ &  (arcsec) & Vortex width & 0.0804\\
                          &  (au) & Vortex width & 8.85\\
$\theta_v$ & (degree)&  Vortex position angle\tablenotemark{3} & -88.6\\
$\sigma_{v,\theta}$ & (degree) & Vortex angular width & 44.6\\
\cutinhead{central inner disk}
$F_g$ & (mJy) & Gaussian flux & 8.50\\
$\sigma_g$ & ($10^{-3}$ arcsec) & Gaussian width & 5.59\\
$\Delta x_g$ & ($10^{-3}$ arcsec) & Offset along de-projected x-axis & -23.1\\
$\Delta y_g$ & ($10^{-3}$ arcsec) & Offset along de-projected y-axis & -7.19\\
\enddata
\tablenotetext{1}{Measured due East from North.}
\tablenotetext{2}{A value of 0$\deg$ is face on, and 90$\deg$ is edge-on.}
\tablenotetext{3}{Measured due East from the system position angle.}
\end{deluxetable*}

The combined vortex and double ring model  allows for a good fit of the image, although the residuals still show some structure, in particular close to the ring inner edge. 
However, none of these two rings or vortex correspond to an outer ring found by \cite{Walsh2014}. 
  We also note that the best fit confirms what is seen by eye: a significant offset between the central Gaussian source and the central position of the ring.

\subsection{Radial cuts}
\begin{figure}[ht]
\plotone{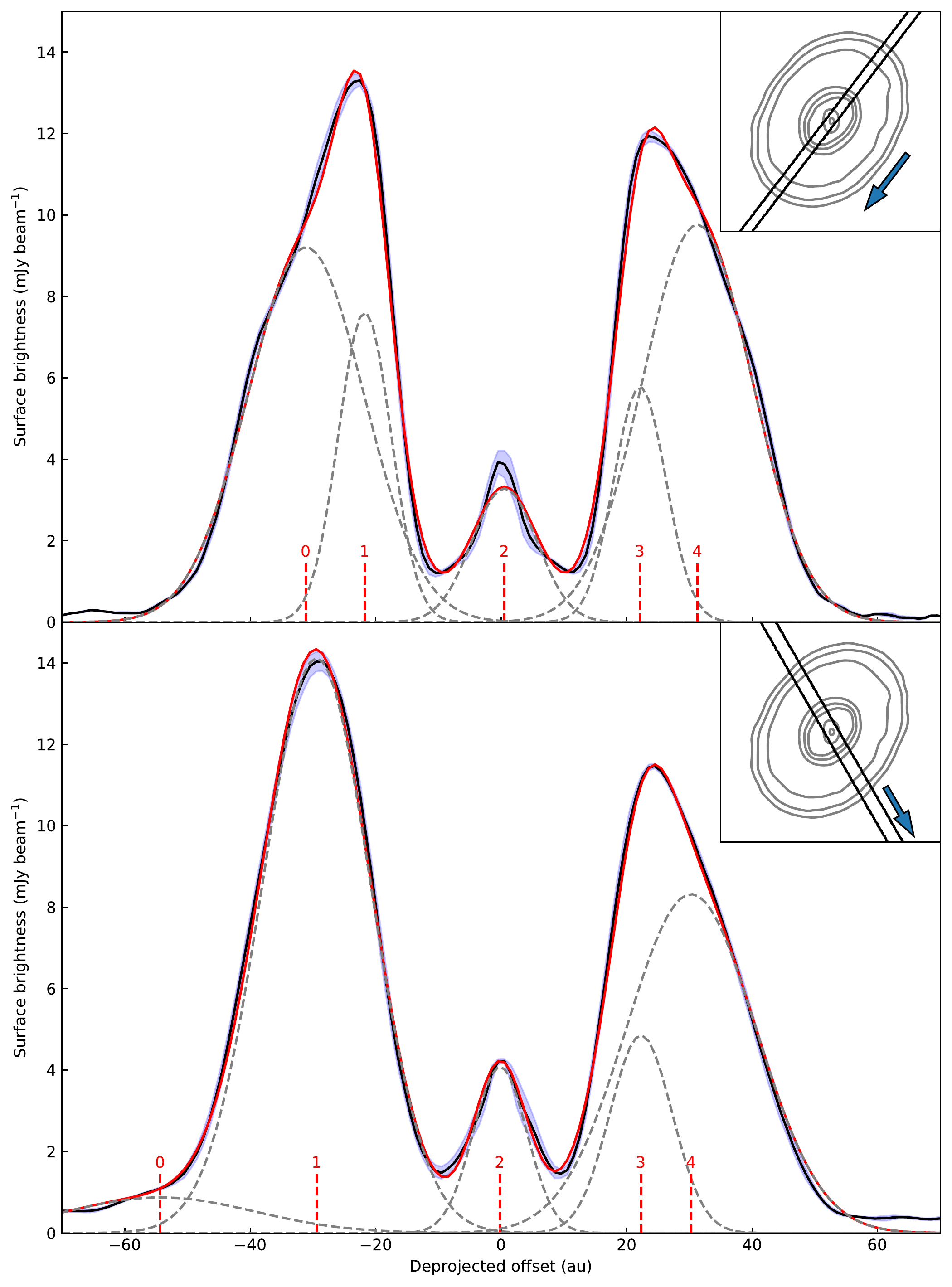}
\caption{The top and bottom panel show radial cuts along the major axis and the vortex, respectively. The black lines are the average of the beam-wide strips. The profiles are clearly asymmetric and  well reproduced by a superposition of 5 Gaussians. The dashed lines show the individual Gaussians (see, Table \ref{table-gaussians}) and the red lines their sum.
An inset showing the orientation of the cuts is presented in the top right corners of each panel. 
\label{fig-cut}}
\end{figure}

We generate two cuts, one through the disk major axis and one through the vortex maximum emission to 
investigate in more detail the ring morphology.
Figure~\ref{fig-cut} shows the average flux along beam-wide strips along both directions. 
The profiles can clearly not be fitted with a single Gaussian flux distribution and they show significant asymmetries in the peak flux on both sides ($\approx 15-25\%$).
Fitting the profiles with a superposition of 5 Gaussians provides a good fit, however. 
The best fit parameters are summarized in Table \ref{table-gaussians}.

\begin{deluxetable}{cCCC}[ht]
\tablecolumns{4}
\tablecaption{Multiple Gaussian fit\tablenotemark{a}\label{table-gaussians}}
\tablehead{
\colhead{Component} & \colhead{Center} & \colhead{Peak flux}& \colhead{$\sigma$}\\
\colhead{} & \colhead{(au)} & \colhead{(mJy beam$^{-1}$)}& \colhead{(au)}
}
\startdata
\cutinhead{NW-SE}
\#0&  -31.1\pm 0.7 &  9.2\pm 0.4 &   8.9\pm 0.3\\
\#1&  -21.7\pm 0.2 &  7.6\pm 0.8 &   3.3\pm 0.4\\
\#2&    0.5\pm 0.3 &  3.3\pm 0.2 &   4.5\pm 0.4\\
\#3&   22.1\pm 0.2 &  5.8\pm 0.7 &   3.2\pm 0.5\\
\#4&   31.3\pm 0.6 &  9.8\pm 0.3 &   8.4\pm 0.2\\
\cutinhead{NE-SW}
\#0&  -54.3\pm 8.7 & 0.88\pm0.09 &  14.7\pm18.5\\
\#1&  -29.4\pm 0.3 & 14.11\pm0.96 &   8.2\pm 0.3\\
\#2&   -0.2\pm 0.3 & 4.08\pm0.08 &   3.4\pm 0.3\\
\#3&   22.3\pm 0.2 & 4.85\pm0.67 &   4.2\pm 0.5\\
\#4&   30.3\pm 0.7 & 8.32\pm0.39 &  10.0\pm 0.2\\
\enddata
\tablenotetext{a}{Each Gaussian is described as:
\[
f(x)=F_{peak}~ {e^{-(x-x_{center})^2/2(\sigma^2+\sigma_{beam}^2)}}
\]}
\end{deluxetable}

We also attempted to fit the profiles with asymmetric Gaussians \citep[see, e.g.,][]{Pinilla2017}, but the results are rather poor and therefore not reported here.

\begin{figure}[h]
\includegraphics[width=0.475\textwidth]{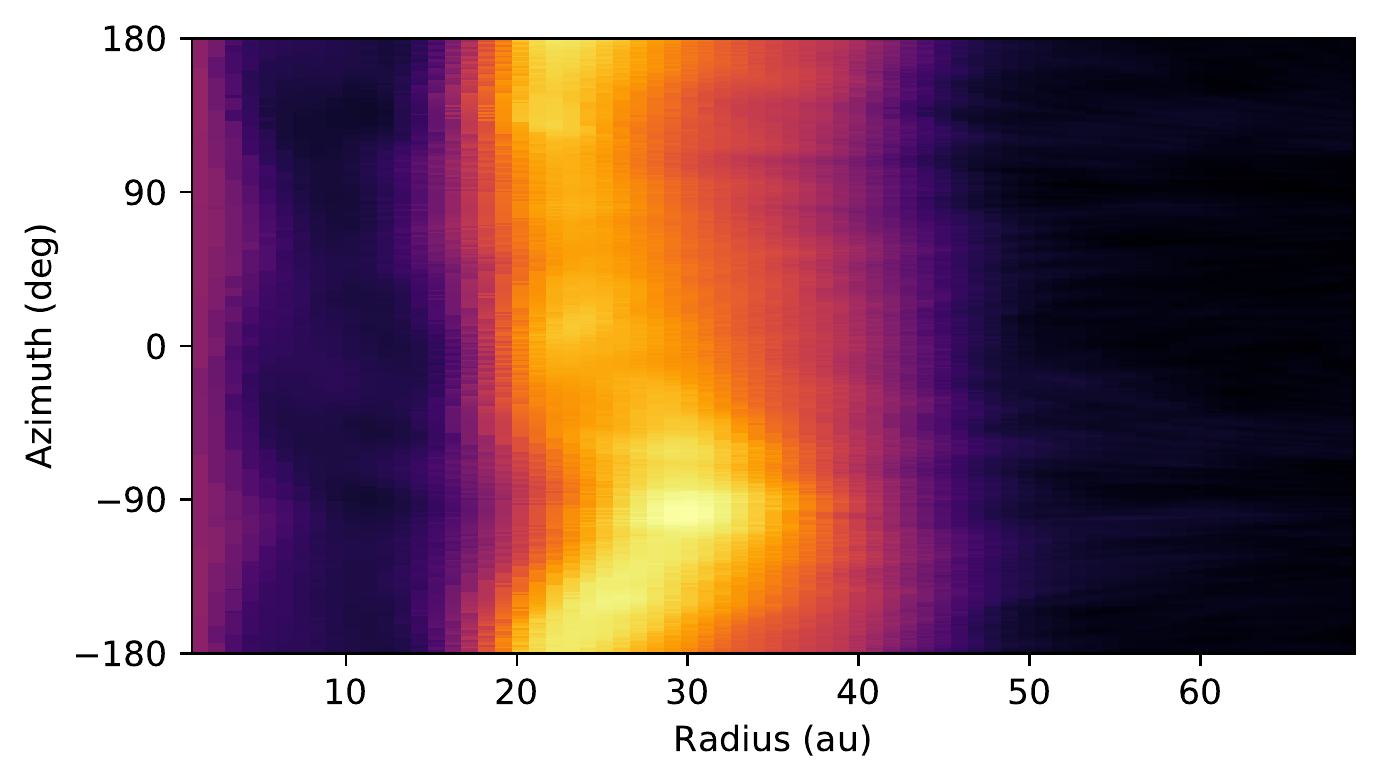}
\caption{\label{fig-polar-map}
Deprojected continuum image in polar coordinates. The ring emission is non-Gaussian, 
as exemplified in Fig.~\ref{fig-cut}. 
The azimuthal angle is measured from North due East from the the disk semi-major axis, with $0\deg$ in the South-East direction.}
\end{figure}

Moreover, we decompose the deprojected map into its polar coordinates (radius and angle) to better 
understand the radial structure of the emission (see Figure.~\ref{fig-polar-map}). 
The plot confirms the radial asymmetry in the main ring, with a ``slow'' flux drop as the radius increases beyond 
the $\approx$30\,au radius. 
On the other hand, the ring emission has a steeper inner edge and clear azimuthal asymmetry. 
However, all these structures are unrelated to a previous outer ring claimed by \cite{Walsh2014} at $\approx 190$\,au.

\subsection{Circumplanetary disk emission\label{sec-cpd-flux}}

Numerical simulations predict the presence of circumplanetary disks around young forming gas giant planets \citep[e.g.,][]{Szulagyi2014_CPD_Accretion,Szulagyi2017_CA_DI,Zhu2016_CPD}. 
The expected disk sizes are supposed to be much smaller than the beam size of 
the observations presented here ($\sim$3\,au). 
Therefore, we expect any CPD emission to appear point-like in our data. 
However, we do not find any evidence for point-like emission close to or around 
the claimed proto-planet positions and place a strong 3-$\sigma$ detection limit 
of 198\,$\mu$Jy for an unresolved source.

\subsection{CO emission}
\begin{figure*}[ht]
\includegraphics[width=\textwidth]{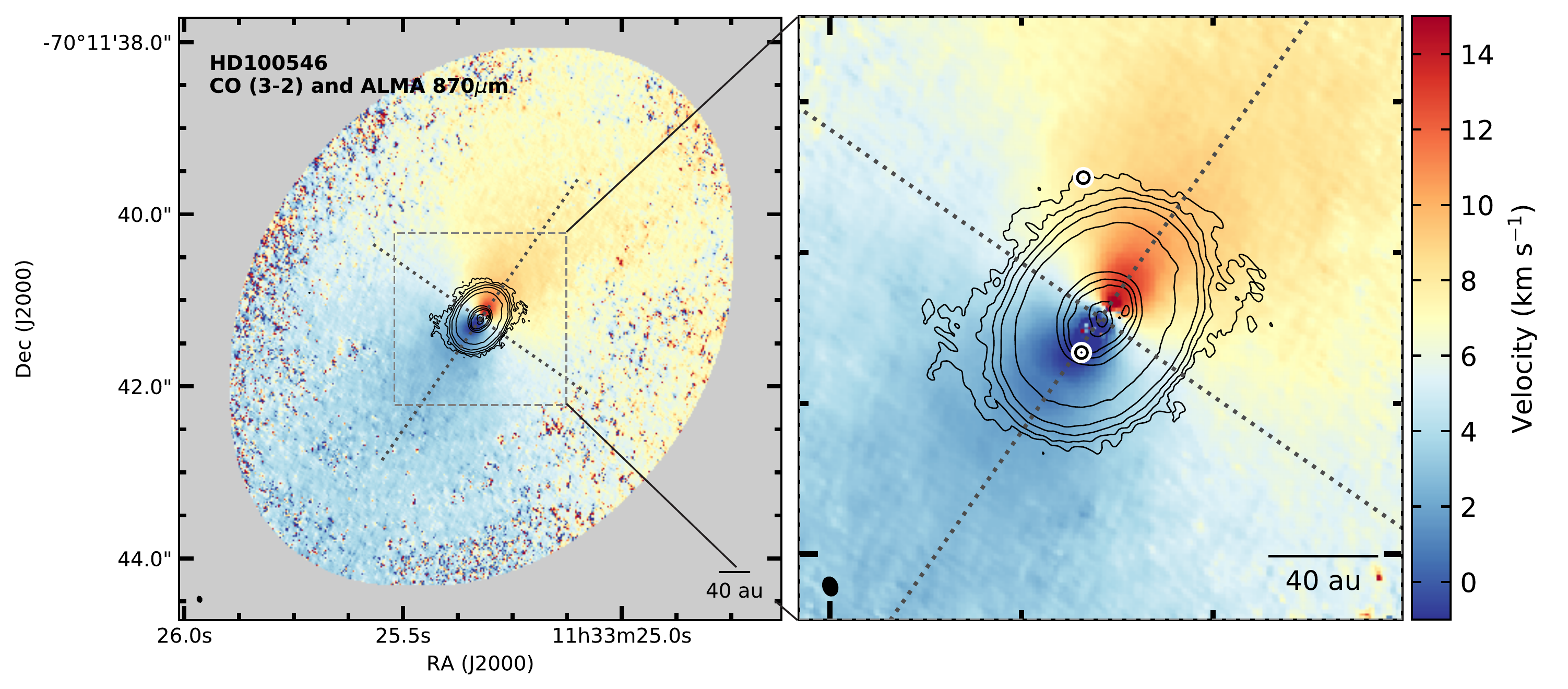}
\caption{CO (3--2) first moment map (centroid velocity) for the HD 100546 disk using the 
Keplerian velocity mask is shown in color. 
The 870\,$\mu$m continuum emission map, using robust briggs weighting, 
is overlaid in contours shown at 
[5, 10, 20, $\ldots$, 320]$\times$rms, where rms is 66\,$\mu$Jy\,beam$^{-1}$. 
Left panel shows the full disk emission, while right panels shows the zoom-in into 
the region of the continuum emission.
Dotted lines show the major and minor axes obtained from fitting the dust continuum visibilities.
Circles show the positions of the two planet candidates for HD 100546. 
The synthetized beam is shown at the bottom left corner.
\label{fig-CO-Vc}}
\end{figure*}

\begin{figure}[ht]
\includegraphics[width=0.5\textwidth]{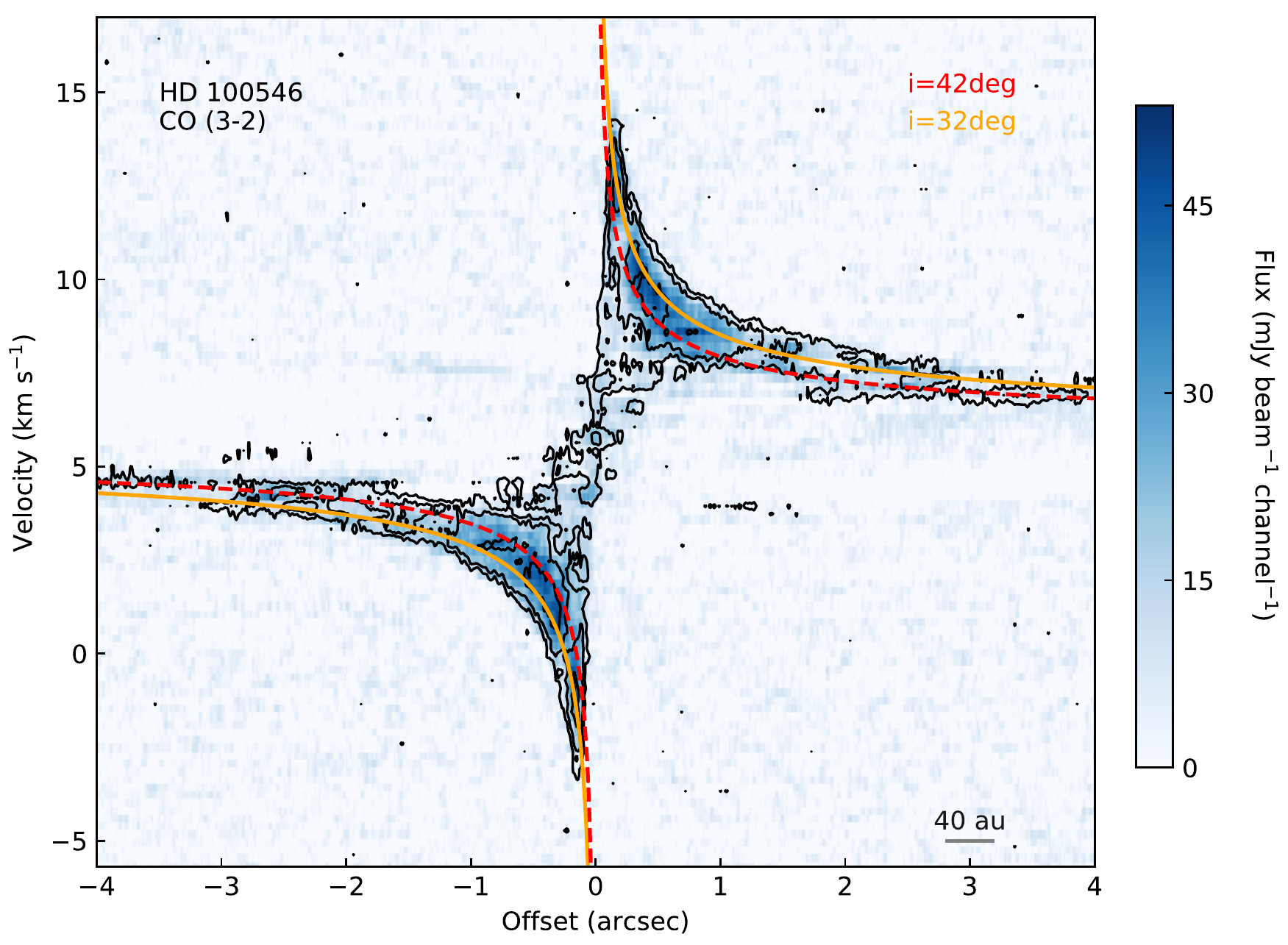}
\caption{PV diagram of CO (3--2) along the major axis shown in Figure~\ref{fig-CO-Vc}.
Contours are shown at [3, 6, 12]$\times$rms, where rms is 3.3\,mJy\,beam$^{-1}$
per channel. 
Orange and red curves show the expected Keplerian velocity for a central star of 2.2\,M$_\odot$ 
and inclination angle of 32$\degr$ and 42$\degr$, respectively.
\label{fig-CO-PV}}
\end{figure}

The CO (3--2) emission (Fig.~\ref{fig-CO-Vc}) extends out to $\approx$2.7\arcsec (300\,au), 
which is less extended than the emission detected with the ALMA Cycle0 data \citep{Pineda2014,Walsh2014} 
because of the missing short baselines in our observations.
Clearly the CO emission is much more extended than the continuum emission, 
which was already identified in the Cycle0 analysis.
The first moment (intensity weighted velocity) map is presented in Figure~\ref{fig-CO-Vc}, 
overlaid with the continuum emission.
The position velocity (PV) diagram along the disk's major
axis is presented in Figure~\ref{fig-CO-PV}. 
The Keplerian velocity profile for the HD 100546 system, with M$_*$=2.2\,M$_\odot$
and 42$\degr$ inclination angle, reproduces the
velocities at a distance$>$2$\arcsec$ from the star (red curve in Figure~\ref{fig-CO-PV}).
For separations $<$2$\arcsec$, the velocities are better reproduced with
an inclination angle  of 32$\degr$ (orange curve in Figure~\ref{fig-CO-PV}).

\section{Discussion\label{sec-discussion}}

\subsection{Circumplanetary disk upper limits}\label{sec:CPDMass}
Given the non-detection of CPD emission towards HD100546 b, 
we place upper limits on the mass or size of the CPD, 
depending on the assumption of optically thin or thick emission.

In the case of the CPD emission being optically thin, 
we estimate the total CPD mass via 
\begin{equation}\label{eq:mass}
M_{total}=\frac{d^2\, F_\nu}{B_\nu(T_{d}) \kappa_\nu\,f_d}~,
\end{equation} 
where $\kappa_\nu$ is the dust opacity per dust mass, $d$ is the distance to the source, $F_\nu$ the observed flux, $B_\nu(T_{d})$ the black body function, and $f_d$ is the dust-to-gas ratio. For the opacity we assume  
 $\kappa_\nu=0.2\,(7\,{\rm mm}~\lambda^{-1})$\,cm$^{2}$\,g$^{-1}$, which is consistent with the value used by \cite{Isella2014-CPD}.
 This opacity assumes a dust composition and grain size distribution as in \cite{Isella2012}. We note that this $\kappa_\nu$ is a factor $\approx$2$\times$ lower than that used by \cite{Beckwith1990,Andrews_2011-Cavities_Disks}, and therefore our mass upper limits are conservative. 
Finally, $f_d$ is  assumed to be 0.01.
Given the emission upper limit determined in section \ref{sec-cpd-flux}, we determine the CPD (dust and gas) mass upper limit in the optically thin case to be 1.44\,$M_{\earth}$.

In the case of the CPD emission being optically thick, the disk radius is calculated from
$
F_\nu=B_\nu(T_d)\,\Omega~,
$
where $T_d$ is the dust temperature of the CPD and $\Omega$ is the area subtended on the sky ($\Omega=\pi R_{CPD}^2/d^2$). 
Therefore, the radius is derived as
\begin{equation}
R_{CPD}= \sqrt{\frac{F_{\nu}}{\pi B_\nu(T_d)}}\,d~.
\end{equation}
An upper limit for the CPD radius of 0.44~au is obtained using a CPD temperature equal to the 
mean dust temperature for the millimeter sized particles in the radiative transfer model 
at that radius ($T_{d,mean}=$53~K, see Sec.\ref{sec-Tprof}), 
while the radius would be only 0.09\,au for a temperature of $932$\,K, 
which is the estimated temperature from high-contrast imaging at L- and M-bands 
\citep{Quanz2015-HD100546b}.
Both numbers are much smaller than the 2.8\,au radius of the Hill sphere expected for a 1\,$M_J$ planet at 53\,au (HD100546 b). 
Several studies have determined the CPD radius to be between 0.3 and 0.5 of the Hill radius 
\citep{Quillen_1998-ProtoJovian_Outflows,Ayliffe_2012-protoplanet_formation,Shabram_2013-Disk_Planet_Simulation}
A conservative CPD radius' upper limit of 0.44\,au  yields an upper limit for the planet mass of 47\,$M_\earth$ (0.15\,$M_J$).

Both cases, optically thin and thick limits, provide important constraints for gas giant planet formation processes by constraining fundamental properties of CPDs. In addition, \cite{Zhu2016_CPD} provided predictions for the SEDs of CPDs including fluxes up 
to the sub-millimeter regime based on the . 
The predicted flux at 870\,$\mu$m is $\sim$800\,$\mu$Jy, which is almost a factor of  10$\times$ the noise level in the image.  On the other hand, detailed numerical simulations and synthetic observations of CPDs 
carried out by \cite{Szulagyi2018-ALMA} show that at large separations from the central star, 
a small fraction of the CPDs ($<R_{Hill}/3$) are warmer than the CSD. 
Furthermore, based on the nominal CSD setup used by \cite{Szulagyi2018-ALMA} 
the expected flux for a CPD around a 1\,M$_J$ planet at 52\,au is $\approx 250\,\mu$Jy, which 
is comparable to the upper limits here reported, and therefore it is still 
consistent with the ALMA observations.

According to \cite{deVal-Borro_2007} even a Neptune-mass planet can generate a vortex of Rossby-Wave Instability, 
so this is consistent with our planetary mass limit. How strong is the vortex is depending on many factors apart 
from the planetary mass: dust-to-gas ratio, viscosity, magnetic field of the disk etc. A detailed parameter study 
of various numerical simulations is needed for this system in order to constrain the planetary mass based on 
the vortex we observe, such as been done for IRS48 \citep{Huang2018}, which is beyond the scope of this work.

\subsection{CPD masses and ages}

\begin{figure}[ht] 
\plotone{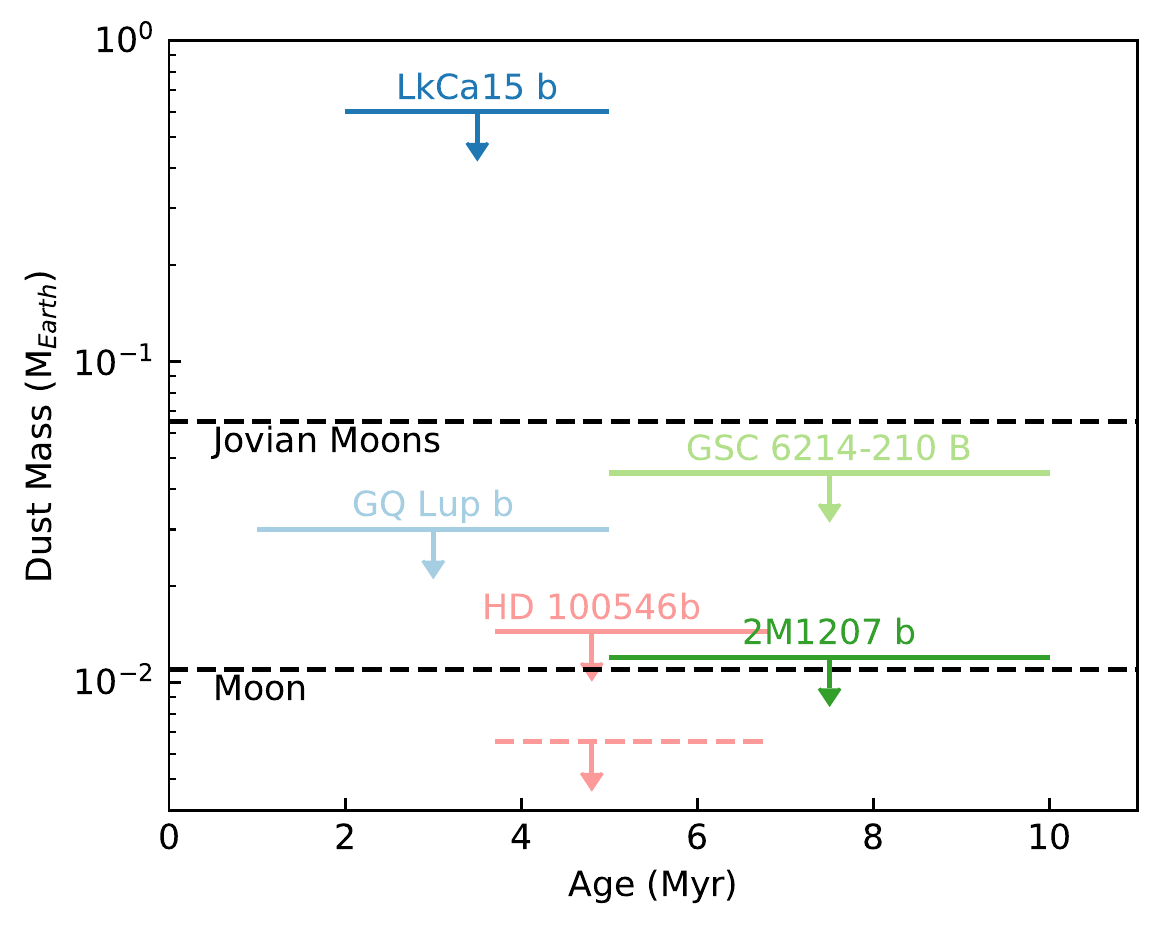}
\caption{Adapted from \cite{Ricci_2017-2M1207}. 
CPD mass upper limits are shown as a function of the central object's estimated age. 
For HD100546 b we show two estimates: (1) the solid red bar shows the value reported in Section~\ref{sec-cpd-flux}, and (2) the CPD mass when using the same dust opacity as for the other CPD estimates shown. We also show the mass contained in the Jovian Moons and the Earth Moon for comparison (dashed lines).
\label{CPD_Mass}}
\end{figure}

Figure~\ref{CPD_Mass} compares the results of a few studies which have provided upper limits for CPD masses  
\citep[see also, ][]{Ricci_2017-2M1207}. 
The CPD mass upper limit obtained for HD100546 b in Sec.~\ref{sec:CPDMass} is comparable 
to that reported by \cite{Ricci_2017-2M1207}, however, our assumed dust opacity is smaller
and therefore, we re-scale the CPD mass estimate to the one used by \cite{Ricci_2017-2M1207} 
and plotted it using dash line in Fig.~\ref{CPD_Mass}. 
This sample includes systems covering a wide range of stellar (host) mass and environments. 
However, it consistently shows that the CPDs, in case they do exist, carry only a small amount of mass.
This is at odds with several models that generate substantial CPDs to feed protoplanets 
\citep{Shabram_2013-Disk_Planet_Simulation,Stamatellos_2015-Disks,Zhu2016_CPD,Gressel_2013-Circumplanetary_Disk}.
On the other hand, the current CPD mass limits are consistent with the ``gas-starved'' disk 
scenario proposed by \cite{Canup2002,Canup2006}, 
as well as the numerical simulations by \cite{Szulagyi2017_CPD_GAP} 
that show a correlation between CPD mass and CSD mass. 

We calculate an upper limit for the potential planet's mass using Eq.~7 from \cite{Szulagyi2017_CPD_GAP}, 
\begin{equation}
M_{CPD}\times10^{4}=3.17\,M_{CSD}\,M_p - 4.33\,M_{CSD}~,
\end{equation}
which relates the CPD, planetary ($M_p$), and CSD mass (all in units of M$_J$).
This assumes that the planet is still accreting from the surrounding CSD, which is supported by the previous detection of L'- and M-band thermal emission.
Assuming the optically thin CPD estimate case from above, 
we place a planetary mass upper limit of 1.65\,$M_J$, using 
our CPD mass upper limit of 1.44\,$M_\earth$  (0.0045\,$M_J$) and the CSD mass of 50\,$M_J$ \citep{Pineda2014}.
This upper limit on the planetary mass estimate is clearly less stringent as the 
one derived using the optically thick approximation in Sec.~\ref{sec:CPDMass}, 
however, the upper limit calculated using the relation between CPD and CSD 
does have a less strong assumption and might be more realistic than the one reported 
in Sec.~\ref{sec:CPDMass}.

\subsection{Central or inner disk}
The central emission is compact and represents the  inner-most circumstellar material. 
We fitted a single Gaussian over a 80\,mas region with a total flux of 
13.6$\pm$1.0\,mJy, a deconvolved FWHM of the major and minor axis of 
80$\pm$8\,mas and 56$\pm$6\,mas, respectively, 
and with a position angle of  177$\pm$14 deg.

We use Eq.~\ref{eq:mass}, the same dust properties used in 
Sec.~\ref{sec:CPDMass}, and a disk temperature of 300~K, to derive 
an inner disk mass of 15\,$M_{\earth}$.
The stellar accretion rate of the central star is estimated to be 
$\dot{M_*} = 10^{-7.04^{+0.13}_{- 0.15}}~{\rm M_\odot\, yr^{-1}}$ 
\citep{Fairlamb2015}.
Thus, the central disk depletion lifetime (${\rm M_{disk}/\dot{M}}$) is only 500~yr. 
Therefore, the disk must be replenished with material from the outer ring/disk 
\citep[e.g.,][]{Pinilla2016}.

\subsection{Comparison with SPHERE scattered light data}
\begin{figure}[ht]
\includegraphics[width=0.5\textwidth]{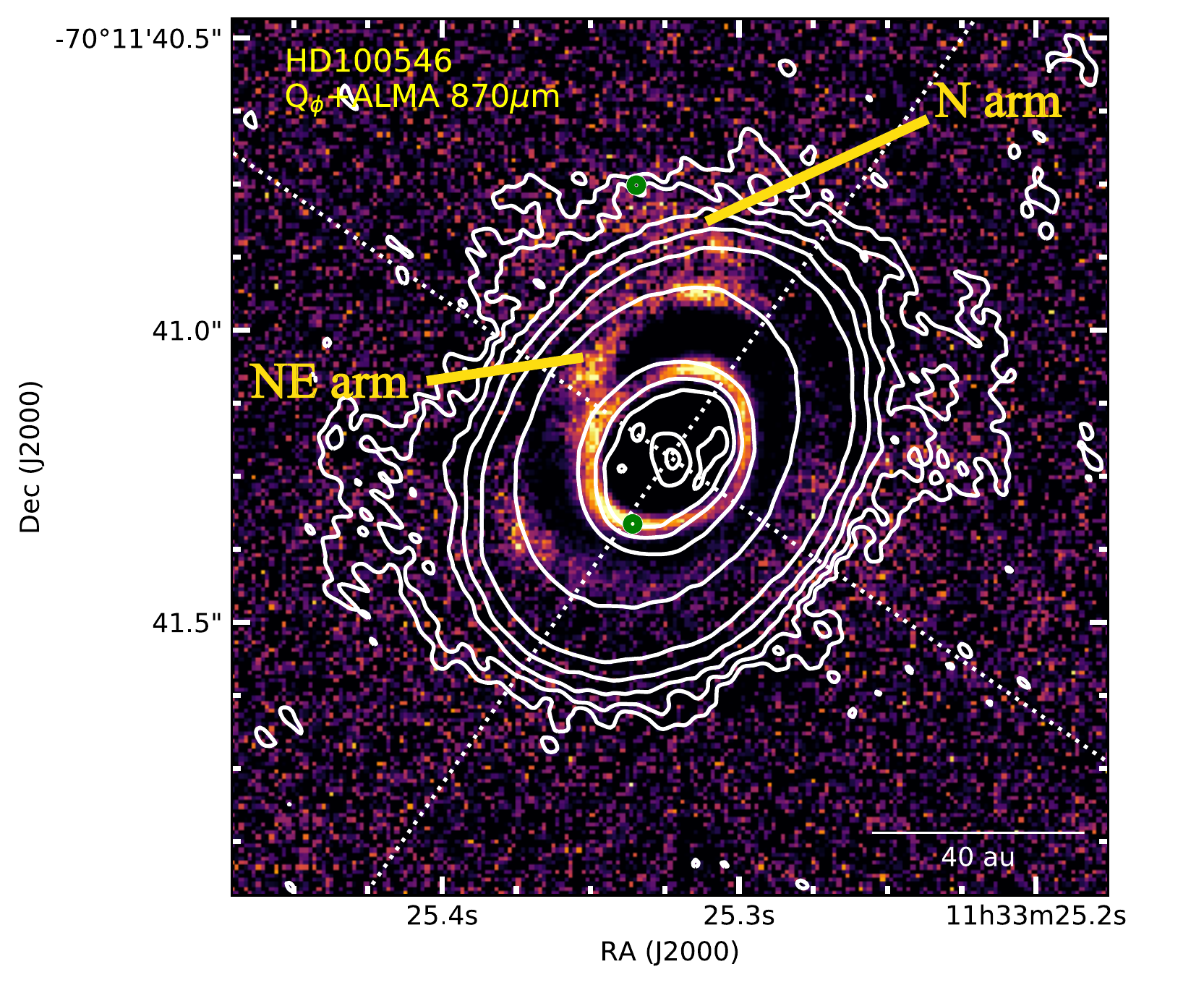}
\caption{Unsharp masked version of a SPHERE/ZIMPOL  $Q_\phi$ image overlaid with our ALMA continuum data (white contours).
Marked are the spiral features identified from the SPHERE data.
The NE-arm feature matches the central location of the ring-like continuum emission.
The N-arm feature is located close to the low-level brightness emission close to HD100546 b.
The green markers show the position of the claimed planets in the system.
\label{fig-ALMA-SPHERE}
}
\end{figure}

\begin{figure*}[ht]
\includegraphics[width=\textwidth]{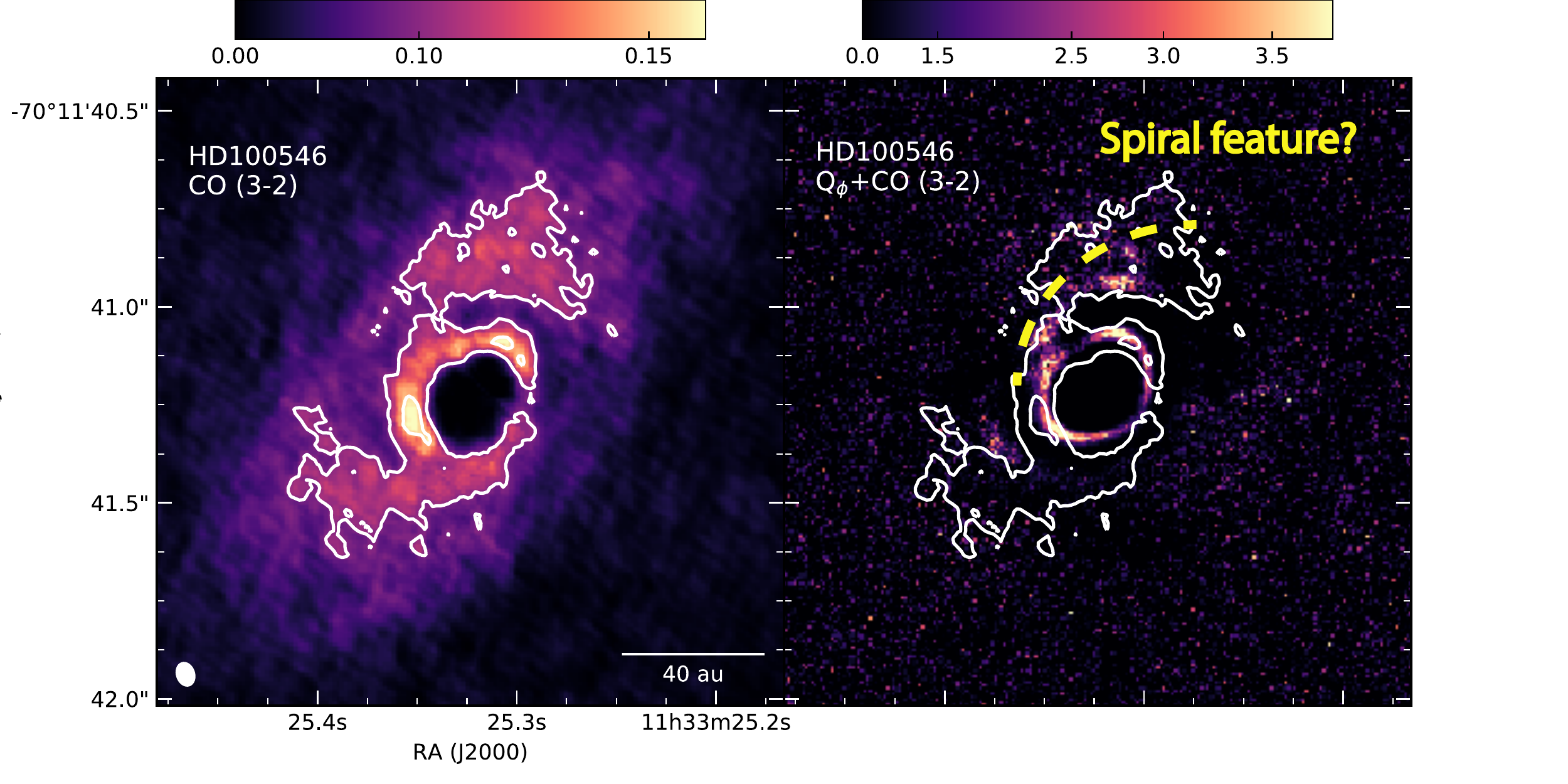}
\caption{Comparison of CO integrated intensity and 
unsharp masked version of a SPHERE/ZIMPOL  $Q_\phi$ image.
\emph{Left:} Background and contours show the 
integrated intensity map of CO (zoomed-in 
version of Fig.~\ref{fig-CO-TdV}) 
using a stretch to highlight the 
suggestive spiral-like emission highlighted by the contours. 
Bottom left and right corner show the beam and scale bar,
respectively.
\emph{Right:} Unsharp masked version of a SPHERE/ZIMPOL 
$Q_\phi$ image (as in Fig.~\ref{fig-ALMA-SPHERE}) 
overlaid with CO integrated intensity contours 
shown in left panel.
The position and general orientation of the NE-arm feature seen with SPHERE 
is similar to the CO enhancement shown in the left panel. 
This would suggest the presence of a real spiral-like feature in the disk surface.
\label{fig-ALMA-SPHERE-CO}
}
\end{figure*}

\cite{Garufi2016} presented an unsharp masked version of the HD100546 disk based on SPHERE/ZIMPOL polarimetric differential imaging data. 
This image shows the disk inner rim, a spiral to the NE, and an arm-like structure to the North. 
In Figure~\ref{fig-ALMA-SPHERE} we show the SPHERE $Q_\phi$ 
image with our ALMA continuum map overlaid in contours, 
while Figure~\ref{fig-ALMA-SPHERE-CO} similarly compares it 
to the CO integrated intensity. 
The SPHERE data are aligned to match the star position with the center of the compact dust continuum emission. 

This comparison confirms that the disk inner rim is well traced by the SPHERE observations and by the ALMA observations 
(continuum and CO). 
The NE-spiral feature observed in the SPHERE data coincides with the central region of the ring in the continuum emission, 
which indicate that the spiral-like feature in scattered light does not have a counterpart in the mid-plane.
However, this feature location and general orientation is coincident with a spiral-like enhancement seen in the CO integrated intensity. 
This coincidence might suggests that the spiral-like feature might be real and present in 
the disk surface. 
This is consistent with the fact that small dust grains and gas are well coupled in those disk regions.

\subsection{Disk kinematics}
Based upon the low-angular resolution CO (3--2) ALMA Cycle0 observations a warp disk was claimed by \cite{Pineda2014}
by comparing the PV diagram along the major axis. 
Since the stellar mass is well constrained, then by over-plotting the expected Keplerian curves it was 
clear that not a single disk inclination could reproduce the observations 
\citep[see][for a similar claim of a warp in AA Tau]{Loomis2017}. 
Further analysis of the same data, but using a more complex modelling tool, also suggested the presence 
of a change in the disk inclination \citep{Walsh2017}. 
Our results also show a similar pattern in the centroid velocity map (Fig.~\ref{fig-CO-Vc}), where the 
inner disk region is slightly twisted. 

The possible disk warp has been suggested before to explain different observations
\citep{Quillen_2006-HD100546_Warp,Panic_2010-HD100546_Kinematics}.
Fig.~\ref{fig-CO-PV} shows the PV diagram along the disk major axis, which shows the 
same behaviour seen from the previous low angular resolution, 
with the kinematics of the outer section of the disk ($>$2\arcsec) being better described by an inclination angle of $\approx $42$\degr$, 
while the inner section of the disk ($<$0.5\arcsec) being better described by an inclination angle closer to $\approx $32$\degr$.
This means that the whole disk is not well described by a single inclination angle.

Also, it has been proposed that departures from the Keplerian velocity field in the disk kinematics 
could provide an independent way to identify the presence of a CPD in HD 100546 \citep{Perez2015-CPD}. 
Unfortunately, the image fidelity and sensitivity of the CO (3--2) data here presented 
do not allow us to identify such a feature. 

\section{Summary\label{sec-summary}}

We presented new ALMA high angular resolution observations of the 870\,$\mu$m dust continuum and 
CO~(3--2) of  HD100546. Our results can be summarized as follows:

\begin{itemize}
\item The ALMA 870\,$\mu$m dust continuum and CO~(3--2) 
observations achieve $\approx$50\,mas resolution, 
and they resolve the disk emission with unprecedented detail.
\item The continuum disk emission is resolved as ring-like (between 20--40\,au) and shows a flux asymmetry 
of $\approx$15--25\%.
\item The disk continuum emission is  well fit by 
two concentric Gaussian rings plus a Gaussian vortex to reproduce the flux asymmetry; this morphology is similar to other disks.
\item Radial cuts show that the disk continuum profile are well fitted using a superposition of multiple Gaussian profiles 
 exemplifying the need for two Gaussian rings to match the two broader and narrower  components of the main ring.
\item We searched for circumplanetary disk (CPD) emission at the location of the embedded planet candidate HD100546 b, but no point-like continuum emission is detected. 
This places strong constraints on the 
CPD mass of 1.44\,$M_\Earth$ and radius of 0.44\,au in the optically thin and thick case, respectively.
\item The CPD mass upper limit is enough to be incompatible with several planet accretion models, while synthetic observations of numerical simulation by \cite{Szulagyi2018-ALMA} provide a CPD flux similar to the upper limit reported here. 
Gas-starved models are also still compatible.
\item We derive an upper limit on the planetary mass of 1.65\,M$_J$ based on a 
numerically calibrated relationship between CSD, CPD, and planetary masses assuming on-going accretion.
\item A central compact emission is also detected, which arises from the inner central disk. 
We estimate an inner disk mass of 15\,$M_\Earth$, and using a previously estimated accretion rate onto the central star, we calculate an inner disk lifetime of 500\,yr.
Therefore, the inner disk must be replenished with material from the outer ring.
\item We compare high angular resolution SPHERE polarization data with ALMA continuum and CO emission. 
This suggest that the NE-arm feature see in the polarized emission does not have a corresponding dust column density feature, 
however, it is well matched by a  spiral-like feature seen as enhanced CO integrated emission.
This is consistent with the expectation the both CO and small dust particles trace the disk surface.
\end{itemize}

\acknowledgments
JEP acknowledges the financial support of the European Research Council (ERC; project PALs 320620).  
JSz acknowledges the support from the Swiss National Science Foundation (SNSF) Ambizione grant PZ00P2\_174115.
SPQ acknowledges the financial support of the SNSF. 
Parts of this work have been carried out within the framework of the National Center for Competence in Research PlanetS supported by the SNSF. 
FM acknowledges support from The Leverhulme Trust, the Isaac Newton Trust and the Royal Society Dorothy Hodgkin Fellowship.
This paper makes use of the following ALMA data: ADS/JAO.ALMA\#2015.1.00806.S. ALMA is a partnership of ESO (representing its member states), NSF (USA) and NINS (Japan), together with NRC (Canada), NSC and ASIAA (Taiwan), and KASI (Republic of Korea), in cooperation with the Republic of Chile. The Joint ALMA Observatory is operated by ESO, AUI/NRAO and NAOJ.
The National Radio Astronomy Observatory is a facility of the National Science Foundation operated under cooperative agreement by Associated Universities, Inc. 
This research made use of \verb-Astropy-, a community-developed core Python package for Astronomy \citep{Robitaille_2013}, 
\verb-matplotlib- \citep{Hunter_2007} and 
APLpy, an open-source plotting package for Python hosted at \url{http://aplpy.github.com}.

\facility{ALMA} 

\software{Astropy \citep{Robitaille_2013}, Matplotlib \citep{Hunter_2007}, CASA \citep{McMullin2007}, 
GALARIO \citep{Tazzari_2017-GALARIO}}

\end{document}